%% file: main.tex
\documentclass[prc,twocolumn,showpacs,nofootinbib,aps,longbibliography,superscriptaddress]{revtex4-1}
\usepackage[utf8]{inputenc}
\usepackage[T1]{fontenc}
\usepackage{color}
\usepackage{amsmath}
\usepackage{url}
\usepackage{graphicx}
\usepackage{amssymb}
\usepackage{appendix}
\usepackage[section]{placeins}
\usepackage{multirow}
\usepackage{dcolumn}
\usepackage{hyperref}
\usepackage{mathrsfs} 
\usepackage{bm}
\usepackage{xcolor}
\usepackage{soul}
\usepackage{array}
\usepackage{lipsum}
\usepackage{relsize}
\usepackage{amsfonts}
\usepackage{comment}
\usepackage{outlines}
\usepackage[normalem]{ulem}
\usepackage{natbib}
\usepackage{booktabs}
\usepackage{mathtools}

\definecolor{linkcolor}{rgb}{0,0,0.40} 
\hypersetup{%
    pdfsubject=Paper,
    pdfkeywords={nuclear physics} {Bayesian} {chiral EFT},
    unicode = true,
    breaklinks = true,
    colorlinks = true,
    linkcolor = linkcolor,
    citecolor = linkcolor,
    menucolor = linkcolor,
    urlcolor = linkcolor
}

\begin{document}

\title{Wavefunction-Based Emulation of Coupled-Channels Scattering with Non-Affinely Parametrized Interactions}

\author{M. Catacora-Rios}
\email{catacor1@msu.edu}
\affiliation{Facility for Rare Isotope Beams, Michigan State University, East Lansing, Michigan 48824, USA}
\affiliation{Department of Physics and Astronomy, Michigan State University, East Lansing, Michigan 48824, USA}

\author{K. Beyer}
\email{beyerk@frib.msu.edu}
\affiliation{Facility for Rare Isotope Beams, Michigan State University, East Lansing, Michigan 48824, USA}
\affiliation{Department of Physics and Astronomy, Michigan State University, East Lansing, Michigan 48824, USA}

\author{P. Giuliani}
\email{giulianp@frib.msu.edu}
\affiliation{Facility for Rare Isotope Beams, Michigan State University, East Lansing, Michigan 48824, USA}

\author{K. Godbey}
\email{godbey@frib.msu.edu}
\affiliation{Facility for Rare Isotope Beams, Michigan State University, East Lansing, Michigan 48824, USA}
\affiliation{Department of Physics and Astronomy, Michigan State University, East Lansing, Michigan 48824, USA}

\author{R. J. Furnstahl}
\email{furnstahl.1@osu.edu}
\affiliation{Department of Physics, The Ohio State University, Columbus, Ohio 43210, USA}

\author{F. M. Nunes}
\email{nunes@frib.msu.edu}
\affiliation{Facility for Rare Isotope Beams, Michigan State University, East Lansing, Michigan 48824, USA}
\affiliation{Department of Physics and Astronomy, Michigan State University, East Lansing, Michigan 48824, USA}

\date{\today}

\begin{abstract}
\begin{description}
\item[Background] Physics based emulators offer a fast and reliable replacement for an exact solution of the scattering problem in nuclear physics.  Previous work  developed a reduced basis emulator for  single-channel scattering using an optical potential to describe elastic scattering. 
\item[Purpose] Since many reactions of interest can be cast as a coupled-channel problem, the purpose of this work is to extend the RBM to a coupled-channel framework.
\item[Method] We generalize the reduced basis method to coupled-channel equations (CC-RBM) to describe inelastic scattering. Although our framework is general, in this work we apply it to reactions where the Hamiltonian coupling term comes from assuming a rotational structure model for the target. From a set of training coupled-channel wavefunctions, we perform a singular value decomposition to obtain a reduced set of basis wavefunctions, and then solve the extended (Petrov-)Galerkin equations in that basis. In addition, the empirical interpolation method is used to expand the potentials. 
\item[Results] We apply the CC-RBM method to elastic and inelastic scattering of neutrons on $^{48}$Ca including a quadrupole coupling to populate the first $2^+$ state, and neutrons on $^{208}$Pb, including an octupole coupling to populate its first $3^-$ state. We demonstrate that the CC-RBM calculated elastic and inelastic cross sections  match those obtained using traditional finite-difference (high-fidelity) methods. We show that the CC-RBM results can reliably reproduce the nuclear scattering cross sections at different energy regimes.
\item[Conclusions] The computational accuracy versus time plots demonstrate that the CC-RBM method efficiently increases precision with increasing basis size. Most importantly, for the precisions required in reaction calculations (a percent on the cross section), we find the CC-RBM method offers roughly one and a half orders of magnitude gain in computational speed compared to the traditional coupled-channels solver. However, we also discuss how this scaling becomes less favorable, the larger the number of channels included in the original coupled-channel set.
\end{description}
\end{abstract}

\maketitle

\section{Introduction}
Nuclear reactions play a fundamental role in understanding the structure and dynamics of atomic nuclei, as well as in describing the processes that govern nucleosynthesis in the universe. Accurate reaction modeling is also essential for applications ranging from nuclear energy and astrophysics to national security and medical isotope production\cite{long_range_2023,Astrophysics_White_paper_2022}. In few-body nuclear reactions, one attempts to capture the relevant degrees of freedom arising from the complex many-body interactions~\cite{potel2017breakup,johnson2020whitepaper}. 
A critical component of few-body reaction theory is the effective interaction between composite particles—the optical potential~\cite{feshbach1958unified}. 



The parameters of optical potentials have been shown to be a major source of uncertainty in the theoretical description of reactions \cite{hebborn2023optical}, and Bayesian studies have sought to quantify, understand, and propagate this parametric uncertainty within various few-body models~\cite{lovell2018,king2019comparison,lovell2020recent,catacora2023complete,catacora2019exploring,catacora2021statistical}. 
However, Bayesian analyses require repeatedly solving the Schr\"odinger equation for many parameter values, a computationally demanding task that has motivated the development of emulators. Emulators based on Reduced Basis Methods (RBMs)~\cite{quarteroni2015reducedpde,bonilla2022training,drischler2023buqeye} can quickly approximate the solution to the full scattering problem within a low-dimensional subspace spanned by representative solutions of the Schr\"odinger equation from a few parameter sets. 

The governing equations for this low-dimensional subspace are obtained by projecting the full scattering operators onto the reduced basis through either the Kohn Variational Principle (KVP)~\cite{kohn1948variational} or a Petrov–Galerkin projection~\cite{quarteroni2015reducedpde}. Such projections ensure that the emulator reproduces the observables of the original scattering problem with orders-of-magnitude speedups and minimal loss of physical interpretability \cite{furnstahl2020efficient,drischler2021toward,drischler2023buqeye,zhang20213body,melendez2021fast,garcia2023wave}.



A class of RBM emulators well adapted for nuclear scattering, the Reduced Order Scattering Emulator (ROSE), was developed and implemented in a user-friendly software in Ref.~\cite{odell2024rose}. 
This approach employs the Empirical Interpolation Method (EIM)
~\cite{barrault2004empirical,grepl2007efficient,quarteroni2015reducedpde}
to efficiently perform scattering calculations with realistic potentials that exhibit non-affine parameter dependence, such as the Woods-Saxon interaction often used in phenomenological optical potentials. 
An additional advantage of ROSE
is that it does not appear to suffer from Kohn anomalies~\cite{Schwarz_anomalies}.

The focus of ROSE was on single-channel scattering, yet the most general few-body reaction problem often results in a set of coupled-channel (CC) equations that need to be solved. The CC framework has been widely used to model nuclear reactions, with applications ranging from elastic (e.g., \cite{cc-elastic}) and inelastic scattering (e.g., \cite{cc-inelastic}), to more complex probes such as transfer \cite{cc-transfer}, breakup \cite{cc-breakup} and fusion \cite{cc-fusion}. Many CC reaction applications involve a large number of channels and significant computation time. Being such an important method in the field of nuclear reactions, it would be extremely useful to develop an efficient and accurate emulator that can replace the full CC calculations, especially for those studies when many reaction calculations are necessary. This work represents the first step in this direction.

In particular, we extend the RBM formalism based on the Galerkin projection used in ROSE to a general 
CC framework.
We apply this framework to neutron–nucleus inelastic scattering using a realistic interaction within a collective 
model for target excitation. 
In doing so, we obtain a general, physics-informed CC emulator capable of treating arbitrary interaction forms with 
non-affine parameter dependencies, providing a unified and efficient tool for emulating complex reaction dynamics.


The remainder of this paper is organized as follows. 
In Sec.~\ref{sec:formalism}, we present a general formulation of the CC equations 
and derive the Petrov--Galerkin equations used to construct the reduced system for this general case, along with the EIM 
employed to efficiently treat non-affinely parametrized interaction potentials. 
Section~\ref{sec:Implementation} outlines an implementation of the emulator and
details the generation of the training and testing data sets, using the {\sc fresco} reaction code \cite{fresco}. 
In Section~\ref{sec:Results} we  
benchmark the emulator 
against coupled-channel calculations obtained using a high-fidelity method (i.e. solutions obtained using traditional finite-difference methods such as Numerov),
analyzing the emulator's accuracy, computational performance, and stability.
Section~\ref{sec:Conclusions} summarizes the main findings and discusses possible extensions of this framework.
The coupling matrix elements and other CC scattering formulas required to compute the observables for  inelastic scattering are provided in Appendix~\ref{app:scattering_details}. 
Finally, a pedagogical derivation of the RBM equations for a two-level system is presented in Appendix~\ref{app:two-state}.

\section{Formalism} \label{sec:formalism}

\subsection{General coupled-channels formalism}\label{Sec:CC_Formalism}

The physical problem under consideration involves quantum mechanical scattering between two composite systems: a projectile and a target, each potentially possessing internal degrees of freedom. 
The total Hamiltonian can be written as the sum of the internal Hamiltonians of the projectile and target and the Hamiltonian describing their relative motion:
\begin{equation}\label{eq:total_hamiltonian}
    H_{\text{tot}} = H_{\text{rel}}(\boldsymbol{r}, \boldsymbol{\xi_t}, \boldsymbol{\xi_p}) + h_{\text{t}}(\boldsymbol{\xi_t}) + h_{\text{p}}(\boldsymbol{\xi_p}),
\end{equation}
where $\boldsymbol{r}$ denotes the relative coordinate between the two bodies and $\boldsymbol{\xi_t}$ and $\boldsymbol{\xi_p}$ are the internal coordinates of the target and projectile systems. Note that the Hamiltonian of the relative motion can be decomposed as $H_\text{rel}(\boldsymbol{r}, \boldsymbol{\xi_t}, \boldsymbol{\xi_p})=T_{\boldsymbol{r}} +  V(\boldsymbol{r}, \boldsymbol{\xi_t}, \boldsymbol{\xi_p})$, where $V(\boldsymbol{r}, \boldsymbol{\xi_t}, \boldsymbol{\xi_p})$ is the scattering potential between projectile and target. The contribution from the center-of-mass motion has been separated and is not included in the following expressions. The eigenstates of the target and projectile satisfy the following eigenequations: 
\begin{equation}
    \begin{aligned}
        h_{\text{t}}(\boldsymbol{\xi_t})\Phi^{\text{t}}_{I_\nu} (\boldsymbol{\xi_t}) = \epsilon_{\nu}^t \Phi^{\text{t}}_{I_\nu} (\boldsymbol{\xi_t}) , \\
        h_{\text{p}}(\boldsymbol{\xi_p})\Phi^{\text{p}}_{s_\nu} (\boldsymbol{\xi_p}) = \epsilon_\nu^p \Phi^{\text{p}}_{s_\nu} (\boldsymbol{\xi_p}).
    \end{aligned}
\end{equation}
Here, the index $\nu$ labels a specific channel, uniquely specified by the internal states of the target and projectile, $I_\nu$ and $s_\nu$. Each channel $\nu$ is therefore associated with a channel energy $E_\nu=E-\epsilon_\nu^t - \epsilon_\nu^p$,  where the beam energy $E$ is reduced by the corresponding projectile and target eigenenergies. The total wavefunction of the scattering system can be expanded in a channel basis as:
\begin{equation}\label{eq:total_wavfunction_decomp}
    \Psi_\lambda = \sum_{\nu=1}^{N_c}|\nu(\boldsymbol{\hat{r}_\nu}, \boldsymbol{\xi_t}, \boldsymbol{\xi_p}) \rangle \frac{1}{r_\nu} \psi_{\nu,\lambda}(r_{\nu}).
\end{equation}
where the ket $\lvert \nu \rangle$ encodes all the quantum numbers necessary to specify the internal structure of the target and projectile as well as their angular-momentum couplings. The sum over $\nu$ is taken over the maximum number of coupled channel states $N_c$ being considered. As an example, using an $l$--$s$ coupling scheme (coupling of the orbital angular momentum and projectile spin), we could have:


\begin{equation}\label{eq:coupling_scheme}
    |\nu(\boldsymbol{\hat{r}_\nu}, \boldsymbol{\xi_t}, \boldsymbol{\xi_p}) \rangle = \bigg [ \big [  i^{l} \boldsymbol{\mathcal{Y}}_{l_\nu}^{m_\nu}(\boldsymbol{\hat{r}_\nu}) \otimes \boldsymbol{\Phi}_{s_\nu}^p(\boldsymbol{\xi_p}) \big ]_j \otimes  \boldsymbol{\Phi}_{I_\nu}^t(\boldsymbol{\xi_t}) \bigg ]_{JM}
\end{equation}
where $\boldsymbol{\mathcal{Y}}_{l_\nu}^{m_\nu}$ is the spherical harmonic representing the angular motion between the projectile and the target. Furthermore, the indices $\nu$ and $\lambda$ in Eq.~\eqref{eq:total_wavfunction_decomp} indicate the asymptotic boundary conditions satisfied by the radial part of the wavefunction. By convention the second index, $\lambda$, is typically taken to be the incoming channel such that asymptotically we have:
\begin{equation}
    \psi_{\nu,\lambda}(r) \xrightarrow[r \to \infty]{} I_{\lambda}(r) - S_{\lambda,\nu} \, O_{\nu}(r),
\end{equation}
where $ S $ is the scattering matrix of the coupled-channel system and $I$($O$) is the incoming (outgoing) free scattering Coulomb function $H^-(\eta, k_\lambda r)$ ($H^+(\eta, k_\nu r)$), and where $\eta$ is the Sommerfeld parameter and $k_\nu$ is the corresponding channel wavenumber (related to the beam energy $E_{\nu}=\frac{\hbar^2 k_{\nu}^2}{2 \mu_{pt}}$, with $\mu_{pt}$ being the reduced mass between projectile and target).  Letting the total Hamiltonian of Eq.~$\eqref{eq:total_hamiltonian}$ act on the total wavefunction defined in Eq.~\eqref{eq:total_wavfunction_decomp} and projecting onto some state $r_\mu \langle \mu(\boldsymbol{\hat{r}_\nu}, \boldsymbol{\xi_t}, \boldsymbol{\xi_p}) |$ we obtain the general set of coupled-channel equations for the radial wavefunctions:
\begin{align}
& \sum_{\nu=1}^{N_c} r_\mu\langle \mu  (\boldsymbol{\hat{r}_\mu}, \boldsymbol{\xi_t}, \boldsymbol{\xi_p})| H_{\text{tot}} - E |  \nu(\boldsymbol{\hat{r}_\nu}, \boldsymbol{\xi_t}, \boldsymbol{\xi_p}) \rangle r_\nu^{-1} \psi_{\nu,\lambda}(r_\nu) 
\\
&= \sum_{\nu=1}^{N_c} N_{\mu \nu} \big[T_\nu(r_\nu) - (E - \epsilon_\nu)\big] \nonumber \psi_{\nu,\lambda}(r_\nu) + \sum_{\nu=1}^{N_c} \hat{V}_{\mu \nu} \psi_{\nu,\lambda}(r_\nu).
\end{align}
Here, the non-orthogonality terms $N_{\mu\nu}$ arise if the two coupled-channels change mass partition. In that case, the scattering potential between projectile and target $\hat{V}$ also becomes:
\begin{equation}
    \begin{aligned}
    N_{\mu\nu} = r_\mu \langle\mu(\boldsymbol{\hat{r}_\mu}, \boldsymbol{\xi_t}, \boldsymbol{\xi_p}) | \nu (\boldsymbol{\hat{r}_\nu}, \boldsymbol{\xi_t}, \boldsymbol{\xi_p}) \rangle r_\nu^{-1}\\
    \hat{V}_{\mu\nu} = r_\mu \langle \mu(\boldsymbol{\hat{r}_\mu}, \boldsymbol{\xi_t}, \boldsymbol{\xi_p}) | V(\boldsymbol{r_\nu}, \boldsymbol{\xi_t}, \boldsymbol{\xi_p})|\nu(\boldsymbol{\hat{r}_\nu}, \boldsymbol{\xi_t}, \boldsymbol{\xi_p}) \rangle r_\nu^{-1}.
\end{aligned}
\end{equation}
Here and for the remainder of the paper we will consider processes in which the partition does not change, which includes elastic and inelastic scattering, allowing us to set $\boldsymbol{r_\mu} = \boldsymbol{r_\nu}$. This leads to the simplified set of coupled channel equations:
\begin{equation}\label{eq:cc_before_expansion}
    \begin{aligned}
        [T_\mu(r)-(E-\epsilon_\mu)]\psi_{\mu,\lambda}(r) = -\sum_{\nu=1}^{N_c} V_{\mu\nu}\psi_{\nu,\lambda}(r) .
    \end{aligned}
\end{equation}
Here the couplings are generated through:
\begin{equation}
    V_{\mu\nu} =  \langle \mu(\boldsymbol{\hat{r}}, \boldsymbol{\xi_t}, \boldsymbol{\xi_p}) | V(\boldsymbol{r}, \boldsymbol{\xi_t}, \boldsymbol{\xi_p}))|\nu(\boldsymbol{\hat{r}}, \boldsymbol{\xi_t}, \boldsymbol{\xi_p}) \rangle, 
\end{equation}
By the symmetries of the system, one can perform a tensor decomposition that separates the coupling potential in a \textit{geometric} part dependent on $\boldsymbol{\xi_t}$, $\boldsymbol{\xi_p}$ and $\boldsymbol{\hat{r}}$, and a scalar function that depends on the relative separation $r$ \cite{thompson2009nuclear}.
In our work, the dependence on any interaction parameters $\boldsymbol{\alpha}$ (e.g. the parameters of a deformed Woods-Saxon), must be included in the \textit{radial} part. Explicitly, this means that we can write the coupling potential as:
\begin{equation}\label{eq:potential_decomposition}
    V(\boldsymbol{r}, \boldsymbol{\xi_t}, \boldsymbol{\xi_p};
    \boldsymbol{\alpha}) = \sum_{\omega=0}^{N_{m}} \mathcal{V}_\omega(\boldsymbol{\hat{r}}, \boldsymbol{\xi_t}, \boldsymbol{\xi_p}) f_\omega (r;\boldsymbol{\alpha}).
\end{equation}
The sum in the equation above runs over the number of terms included in the tensor decomposition of the potential up to $N_m$. Throughout the remainder of this work, we will use a semicolon to distinguish between coordinate arguments and parametric dependencies in the potential functions. The tensor decomposition of the coupling potential could be, for example, a multipole expansion of the deformed potential as is frequently adopted in collective model descriptions of nuclear inelastic scattering, where target excitations are modeled as deformations or vibrational modes \cite{temura}. Expanding Eq.~\eqref{eq:cc_before_expansion} using Eq.~\eqref{eq:potential_decomposition} we get the final form of the coupled-channel equations:

\begin{align}
        [T_\mu(r)-(E-\null&\epsilon_\mu)]\psi_{\mu,\lambda}(r) \notag\\ 
        &= -\sum_{\nu=1}^{N_c} \sum_{\omega=0}^{N_m} \mathcal{V}_{\omega\mu\nu} f_\omega (r;\boldsymbol{\alpha}) \psi_{\nu,\lambda}(r),
        \label{eq:cc_final}
\end{align}
where
\begin{equation}\label{eq:couplings_term}
    \mathcal{V}_{\omega\mu\nu}=\langle\mu (\boldsymbol{\hat{r}}, \boldsymbol{\xi_t}, \boldsymbol{\xi_p}) |\mathcal{V}_{\omega}(\boldsymbol{\hat{r}}, \boldsymbol{\xi_t}, \boldsymbol{\xi_p})|\nu(\boldsymbol{\hat{r}}, \boldsymbol{\xi_t}, \boldsymbol{\xi_p}) \rangle.
\end{equation}
The \textit{geometric} term of the coupling potential, $\mathcal{V}_{\omega\mu\nu}$, now acts purely as a scalar coupling strength which depends only on the internal coordinates $\boldsymbol{\xi_t}$, $\boldsymbol{\xi_p}$ of the two subsystems and the relative angular coordinate $\boldsymbol{\hat{r}}$. The component index $\nu$ on the right-hand side of Eq.~\eqref{eq:cc_final}, runs over all the allowed couplings between $\nu$ and $\lambda$, $\nu=1, 2, 3, .., N_c$. The choice of the coupling potential as well as the choice of internal eigenstates $\Phi$ will determine the number of couplings, as can be seen in Eq.~\eqref{eq:couplings_term}. In general, there will be a finite set of couplings, for example using the $l-s$ coupling scheme shown in Eq.~\eqref{eq:coupling_scheme}, angular momentum rules limit the couplings to states within a given $J$ and a given total parity $\pi$:
\[   \pi=(-1)^l\pi^p \pi^t.     \]
 Here $\pi_p$, and $\pi_t$ are the intrinsic parities of the projectile and the target and $l$ is the orbital angular momentum. Therefore, in a general scattering problem, the coupled-channel equations will be solved per total angular momentum number $J$ and total parity. 
\begin{align}
    \label{eq:cc_final_per_J}
        [T_\mu(r)-(E-\null&\epsilon_\mu)]\psi_{\mu,\lambda}^{J^\pi}(r) \notag\\ 
        &= -\sum_{\nu=1}^{N_c} \sum_{\omega=0}^{N_m} \mathcal{V}_{\omega\mu\nu}^{J^\pi} f_\omega (r;\boldsymbol{\alpha}) \psi_{\nu,\lambda}^{J^\pi}(r).
\end{align}
For notational simplicity, we will omit the superscript $J^\pi$, but we emphasize that all the coupled equations in the following sections must be solved for each coupled $J^\pi$ block unless otherwise stated. We also note that the radial form-factor $f_\omega$ no longer carries any dependence on the quantum numbers of the projectile-target system.

\subsection{The Reduced Basis Method for coupled channels}
\label{Sec: RBM}

\subsubsection{Training space}
The Reduced Basis Method (RBM) for neutron-nucleus elastic scattering, as developed in the ROSE framework~\cite{odell2024rose}, can be extended to the general class of coupled channel equations shown in Eq.~\eqref{eq:cc_final}. We denote \(\psi_{\nu,\lambda}(r; \boldsymbol{\alpha})\) as the radial wavefunction in channel $\nu$, resulting from an incoming wave in channel $\lambda$, under an interaction parametrized by $\boldsymbol{\alpha}$ (e.g., Woods-Saxon potential parameters).
The key component in constructing a coupled channels emulator is to identify a suitable low-dimensional subspace onto which high-fidelity solutions for each channel can be accurately projected. 
Once such a subspace is determined, the full high-fidelity wavefunction $\psi$ can be approximated by $\hat{\psi}$ as follows:
\begin{align}
    \psi_{\nu,\lambda}(r;\boldsymbol{\alpha}) &\approx \hat{\psi}_{\nu,\lambda}(r;\boldsymbol{\alpha})  
    \notag \\
    &= \phi_{\nu} (r)\delta_{\nu\lambda} + \sum_{k=1}^{N_{\psi}} c^{(k)}_{\nu}(\boldsymbol{\alpha}) \tilde{\psi}^{(k)}_{\nu,\lambda}(r).
    \label{eq:linear_approx}
\end{align}
Here, the expansion in the reduced basis elements $\tilde{\psi}$ runs up to $N_{\psi}$, the number of basis states chosen in the approximation. The first term on the right-hand side, $\phi_{\nu}$, represents the free solution—i.e., the solution in the absence of any nuclear interaction. A standard Kronecker delta is used to indicate that this term contributes only when the outgoing channel $\nu$ matches each incoming channel $\lambda$. We will call the channel for which $\nu=\lambda$ the elastic channel, and those for which $\nu\neq\lambda$ the inelastic channels. The dependence of the expansion coefficients $c^{(i)}_{\nu}(\boldsymbol{\alpha})$ on the potential parameters $\boldsymbol{\alpha}$ has been made explicit. We note that the expansion of the approximate wavefunction is made component by component, this means that each $\nu=1,2,..,N_c$ in a $J_\pi$ coupled block, each incoming channel $\lambda$ will have its own expansion. The chosen basis functions $\tilde{\psi}^{(k)}_{\nu,\lambda}(r)$, will also be channel dependent, that is: $\tilde{\psi}^{(k)}_{\nu,\lambda}\neq\tilde{\psi}^{(k)}_{\mu,\lambda}$ for $\mu\neq\nu$. This channel-dependent choice of wavefunctions ensures that the asymptotic boundary conditions of the emulated wavefunctions are satisfied by construction.

We will now follow a similar procedure to the ROSE framework~$\cite{odell2024rose}$ to get an orthonormal set of basis functions that define a subspace of solutions. We will choose a basis $\{ \tilde{\psi}^{(k)}_{\nu,\lambda}(r) \}$ by performing a principal component analysis (PCA), also known as proper orthogonal decomposition \cite{quarteroni2015reducedpde}. The PCA is performed on the space spanned by $N_s$ \textit{snapshots}, that is, $N_s$ high-fidelity solutions $\{ \psi_{\nu,\lambda}(r;\boldsymbol{\alpha_i}) \}_{i=1}^{N_s}$ to Eq.~\eqref{eq:cc_final}, generated from the set of parameters $\{ \boldsymbol{\alpha_i} \}_{i=1}^{N_s}$. This gives rise to the notion of a \textit{training space} for the emulator. For those wavefunctions where we have that $\nu=\lambda$, the \textit{snapshots} will be modified by subtracting the free solution. Our $N_{\psi}$ basis functions are then the first $N_{\psi}$ principal components defined as:
\begin{equation}
    \{ \tilde{\psi}^{(k)}_{\nu,\lambda}(r) \}_{k=1}^{N_\psi} = \text{PCA}\bigg [ \{  \psi_{\nu,\lambda}(r;\boldsymbol{\alpha_i}) - \phi_{\nu}(r)\delta_{\nu\lambda}    \}_{i=1}^{N_s}\bigg] .
\end{equation}

\begin{figure}[t!]
    \centering
    \includegraphics[width=1.0\linewidth]{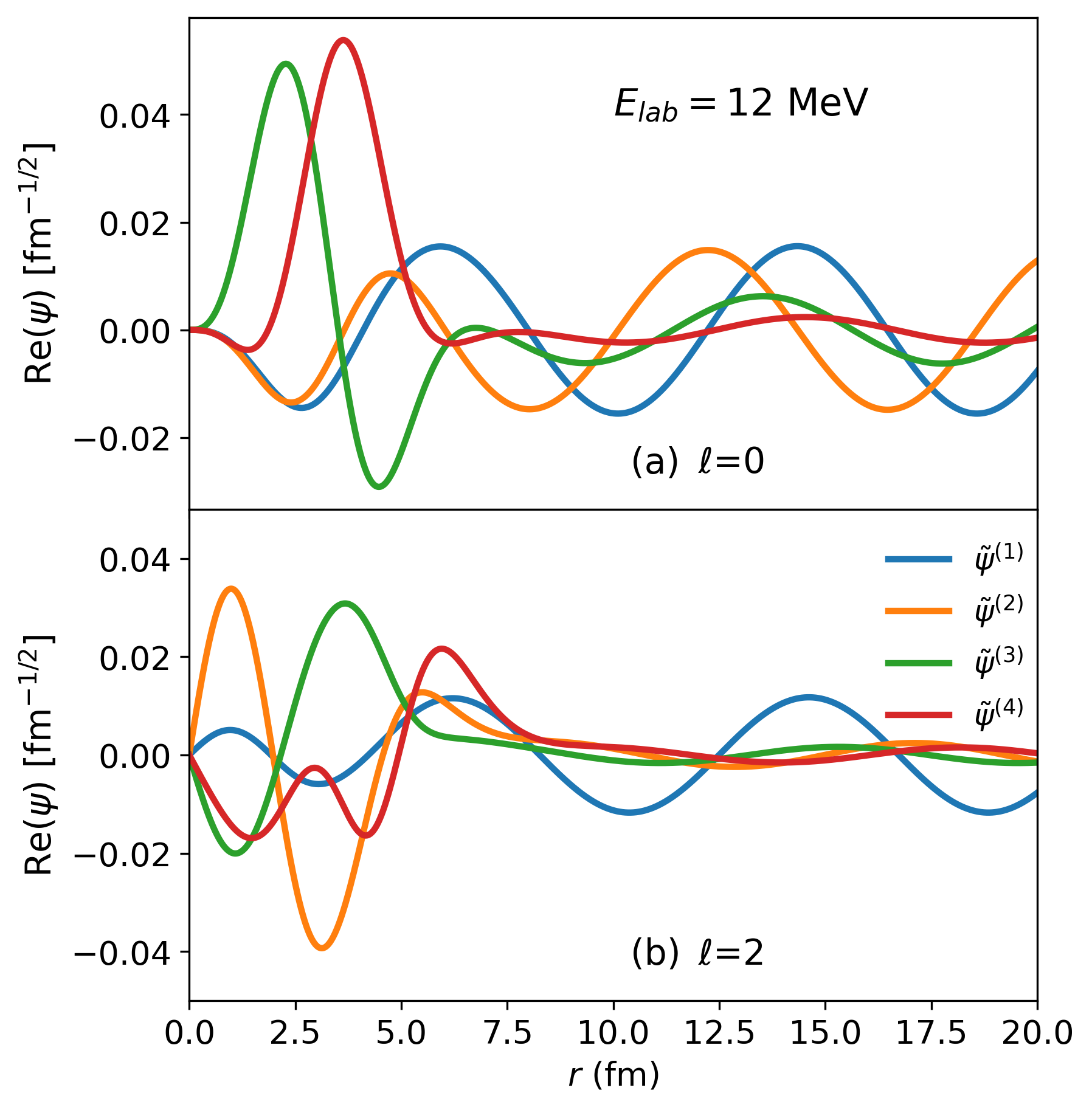}
    \caption{Real parts of the first four PCA components for the two coupled channels in the $^{48}$Ca$(n,n')$$^{48}$Ca$(2^+)$ system at $E_{\mathrm{lab}} = 12$ MeV and $J^\pi = 0^+$.
Panel (a) shows the $l = 0$, $I_t = 0^+$ channel, and panel (b) the $l = 2$, $I_t = 2^+$ channel.}
    \label{fig:SVD_components}
\end{figure}

Naturally we have the requirement that $N_s \geq N_\psi$. This choice of basis elements can be understood as the $N_\psi$ most relevant directions of variability in the \textit{snapshots}, and in the channel where $\nu=\lambda$, the variability is defined with respect to the solution without a nuclear potential. 
Figure~\ref{fig:SVD_components} shows 
the first four PCA vectors obtained for a realistic $^{48}\text{Ca}(n,n')^{48}\text{Ca}(2^+)$ at $E_\text{lab}=12$ MeV calculation.
\subsubsection{The Petrov-Galerkin equations}

After defining the reduced basis, the next step is to formulate a system of equations to determine the expansion coefficients $c_{\nu}^{(k)}(\boldsymbol{\alpha})$ in the approximate wavefunction. Several frameworks have been developed to achieve this in the context of scattering (see e.g. \cite{garcia2023wavefunctionmomentum,zhang20213body}). For this work, we adopt the Petrov-Galerkin approach, a generalization of the Galerkin method used in the ROSE framework~\cite{odell2024rose}. In this method the coefficients are found by projecting the radial part of the Hamiltonian acting on the approximate solution onto the subspace spanned by some functions $\rho^{(j)}_{\nu,\lambda}(r)$, for $j=1,2,..,N_\psi$. As previously stated, each channel component $\nu$ of the wavefunction defined in Eq.~\eqref{eq:cc_final}, must have its own RBM approximation using Eq.~\eqref{eq:linear_approx}. We choose the case where $\rho^{(j)}_{\nu,\lambda}=(\tilde{\psi}^{(k)}_{\nu,\lambda})^*$, where $^*$ denotes complex conjugation. Therefore, the before-mentioned projection leads to the following set of coupled equations:
\begin{equation}
    \begin{aligned}\label{eq:petrov-galerkin}
        \sum_{k=1}^{N_\psi} \bigg[ c_{\mu}^{(k)}\mathcal{D}^{(jk)}_{\mu,\lambda} & + \sum_{\nu=1}^{N_c} c_{\nu}^{(k)}\mathcal{U}^{(jk)}_{\mu\nu,\lambda}(\boldsymbol{\alpha)}   \bigg] \\
        & \quad \quad \quad + \mathcal{R}^{(j)}_{\mu,\lambda} + \sum_{\nu=1}^{N_c} \mathcal{C}^{(j)}_{\mu\nu,\lambda}(\boldsymbol{\alpha}) = 0 \\
        & \quad \quad \quad \quad \quad \quad \text{for $\mu = 1,2,..,N_c$},
    \end{aligned}
\end{equation}
where we have defined:
\begin{equation}\label{eq:petrov-galerkin-def}
    \begin{aligned}
     \mathcal{D}^{(jk)}_{\mu,\lambda} &= \langle \tilde{\psi}_{\mu,\lambda}^{(j)} | T_\mu (r)- (E-\epsilon_\mu ) | \tilde{\psi}_{\mu,\lambda}^{(k)}\rangle, \\
     \mathcal{U}_{\mu\nu,\lambda}^{(jk)}(\boldsymbol{\alpha}) &= \sum_{\omega=0}^{N_m} \mathcal{V}_{\omega\mu\nu}\langle \tilde{\psi}_{\mu,\lambda}^{(j)} |f_\omega(r; \boldsymbol{\alpha})|\tilde{\psi}_{\nu,\lambda}^{(k)} \rangle,\\
     \mathcal{R}_{\mu,\lambda}^{(j)} &= \langle \tilde{\psi}_{\mu,\lambda}^{(j)} |T_\mu (r)- (E-\epsilon_\mu ) | \phi_{\mu}\rangle\delta_{\mu\lambda}, \\
     \mathcal{C}_{\mu\nu,\lambda}^{(j)} (\boldsymbol{\alpha})  &=  \sum_{\omega=0}^{N_m} \mathcal{V}_{\omega\mu\nu}\ \langle \tilde{\psi}_{\mu,\lambda}^{(j)} | f_\omega (r; \boldsymbol{\alpha})| \phi_{\nu}\rangle\delta_{\nu\lambda}.
    \end{aligned}
\end{equation}
 Here the bra-ket notation denotes integration over the radial coordinate $r$. Also note that with the above choice of $\rho_{\nu,\lambda}^{(j)}$ we have $ \langle \tilde{\psi}_{\mu,\lambda}^{(j)} | =|\tilde{\psi}_{\mu,\lambda}^{(j)}\rangle$. These equations can be cast into an algebraic linear system for the desired set of coefficients $ \{ c^{(k)}_{\mu} \}_{\mu=1}^{N_c}$, which are obtained by solving Eq.~\eqref{eq:petrov-galerkin} for all the channels in a $J_\pi$ block simultaneously. These equations are worked out in detail, in their matrix form, for a two-level system in Appendix~\ref{app:two-state}.

\subsubsection{The Empirical Interpolation method for deformed potentials}
When building an emulator, it is useful to conceptualize the computation as occurring in two distinct stages: an \textit{offline stage}, which consists of calculating and extracing $N_s$ snapshots from the high-fidelity solver, which comprise the training space $\{ \psi_{\nu,\lambda}(r;\boldsymbol{\alpha_i}) \}_{i=1}^{N_s}$, followed by performing the PCA and then pre-computating of the matrices Eq. \,\eqref{eq:petrov-galerkin-def}.  This enables the \textit{online stage}, in which one desires to quickly emulate the solution for a new set of parameters $\boldsymbol{\alpha}$. This can be done efficiently only if operations in the full, high-dimensional space used by the high-fidelity solver can be avoided, e.g. by pre-computing the matrix elements of the matrices in Eq.\,\eqref{eq:petrov-galerkin-def}. This would be straightforward if the parameters $\boldsymbol{\alpha}$ were \textit{affine}—that is, if the radial part of the potential could be expressed in a separable form:
\begin{equation}
    f_\omega(r,\boldsymbol{\alpha}) = \sum_i g_\omega^{(i)}(\boldsymbol{\alpha}) u^{(i)}_\omega(r),
\end{equation}
where $g_\omega^{(i)}$ are general functions of the full vector of parameters $\boldsymbol{\alpha}$
and $u_\omega^{(i)}(r)$ are the corresponding radial functions. Such an affine form of the potential would allow all integrals in Eq.~\eqref{eq:petrov-galerkin-def} to be precomputed during the \textit{offline stage}, so that the \textit{online} computation reduces to simple weighted sums involving the parameters $\boldsymbol{\alpha}$. However, most nucleon-nucleus scattering potentials do not allow for such a factorized representation for all parameters. Therefore, we will now use the Empirical Interpolation Method (EIM)~\cite{quarteroni2015reducedpde,hesthaven2016certified,barrault2004empirical,grepl2007efficient} to make these calculations affine. To do this, we will approximate each \( f_\omega(r;\boldsymbol{\alpha}) \) in Eq.~\eqref{eq:cc_final} as a linear combination of interaction basis functions, weighted by parameter-dependent coefficients:
\begin{equation}f_\omega(r;\boldsymbol{\alpha}) \approx \sum_{i=1}^{N_{\text{U}}} b_{\omega}^{(i)}(\boldsymbol{\alpha}) u^{(i)}_\omega(r),
\end{equation}
where \( \{ u^{(i)}_\omega(r) \}_{i=1}^{N_{\text{U}}} \) are a reduced basis for the radial interaction functions $f_\omega$,
and \( b^{(i)}_\omega(\boldsymbol{\alpha}) \) are the corresponding interpolation coefficients encoding the dependence on the interaction parameters \( \boldsymbol{\alpha} \). We obtain these basis functions by sampling the parameter space in a similar manner as for the reduced basis elements. We calculate a set of interactions $\{ f_\omega (r,\boldsymbol{\alpha_j}) \}_{j=1}^{n_U}$ and then retain the $N_U$ most important components of a PCA analysis (note that we once again require $n_U \geq N_U$):
\begin{equation}
   \{ u_\omega^{(i)}(r)\}_{i=1}^{N_U} = \text{PCA} \big [ \{ f_\omega (r,\boldsymbol{\alpha_j}) \}_{j=1}^{n_U} \big ].
\end{equation}
Then the equation that determines the coefficients can be expressed as:
\begin{equation}
    \begin{aligned}
        f_\omega(r_j,\boldsymbol{\alpha})-\sum_{i=1}^{N_U}b_\omega^{(i)}(\boldsymbol{\alpha})u_\omega^{(i)}(r_j) &= 0 \\
        \quad &\text{for $j=1,2,..,N_U$}.
    \end{aligned}
\end{equation}
This equation can be understood as interpolating the value of the radial interaction $f_\omega(r,\boldsymbol{\alpha})$ at $r_j$ to determine its values at all other points $r$. The MaxVol algorithm~\cite{Goreinov_2010} we use to select the radial points $r_j$ is described in the ROSE framework \cite{odell2024rose}, and it consists of maximizing the determinant of a sub-matrix containing snapshots of the interactions over the coordinate grid. This system of equations can then be cast into a linear system involving a $N_U\times N_U$ matrix. Explicitly, we have:
\begin{equation}\label{eq:EIM_coefficients}
    \boldsymbol{b}_\omega(\boldsymbol{\alpha}) = (\boldsymbol{U^{\text{EIM}}}_\omega)^{-1}\cdot \boldsymbol{c^{\text{EIM}}}_\omega(\boldsymbol{\alpha}),
\end{equation}
with
\begin{equation}
    \label{eq:EIM_matrix_form_a}
        \boldsymbol{U^{\text{EIM}}}_\omega =  \begin{bmatrix}
        u^{(1)}_\omega(r_1) & u^{(2)}_\omega(r_1)  & ...  & u^{(n_U)}_\omega(r_1)  \\
        u^{(1)}_\omega(r_2) & u^{(2)}_\omega(r_2)  & ...  & u^{(n_U)}_\omega(r_2)  \\
        \vdots  & \vdots  & \vdots  & \vdots  \\
        u^{(1)}_\omega(r_{N_U}) & u^{(2)}_\omega(r_{N_U})  & ...  & u^{(n_U)}_\omega(r_{N_U}) 
    \end{bmatrix}_{N_U\times N_U} 
\end{equation} 
and
\begin{equation}    
\label{eq:EIM_matrix_form}\boldsymbol{c}^{\text{EIM}}_\omega(\boldsymbol{\alpha}) = \begin{bmatrix}
    f_\omega(r_{1},\boldsymbol{\alpha}) \\
    f_\omega(r_{2},\boldsymbol{\alpha}) \\
            \vdots \\
    f_\omega (r_{N_U},\boldsymbol{\alpha})
        \end{bmatrix}.
\end{equation}
Since the matrix $\boldsymbol{U}_\omega^{\text{EIM}}$ is independent of the potential parameters $\boldsymbol{\alpha}$ one can invert this matrix in the \textit{offline stage} of the calculation and for each set of new parameters $\boldsymbol{\alpha}$, simply evaluate the potential at the corresponding locations $r_j$ and multiply as shown in Eq.~\eqref{eq:EIM_coefficients}. Therefore, using the EIM we are able to perform all radial integrations in \eqref{eq:petrov-galerkin-def} only in the \textit{offline stage} of the emulator, using the fixed basis states. We can then expand the coupling potential as:
\begin{equation}
    \begin{aligned}\label{eq:EIM_coupling_potential_expansion}
        V_{\mu\nu}(r,\boldsymbol{\alpha}) &= \sum_{\omega=0}^{N_m} \mathcal{V}_{\omega\mu\nu} f_\omega(r;\boldsymbol{\alpha}) \\
        &\approx \sum_{\omega=0}^{N_m} \sum_{i=1}^{N_U} \mathcal{V}_{\omega\mu\nu}b_\omega^{(i)}(\boldsymbol{\alpha})u_\omega^{(i)}(r),
    \end{aligned}
\end{equation}
where $\mathcal{V}$ is the \textit{geometric} coupling matrix element defined in Eq.~\eqref{eq:couplings_term}.
With this definition we can update Eq.~\eqref{eq:petrov-galerkin}:
\begin{equation}\label{eq:update_eim_coeffs}
    \begin{aligned}
\mathcal{U}_{\mu\nu,\lambda}^{(jk)}(\boldsymbol{\alpha}) &= \sum_{\omega=0}^{N_m} \sum_{i=1}^{N_U} \mathcal{V}_{\omega\mu\nu}b_\omega^{(i)}(\boldsymbol{\alpha})\langle \tilde{\psi}_{\mu,\lambda}^{(j)} |u^{(i)}_\omega(r)|\tilde{\psi}_{\nu,\lambda}^{(k)} \rangle\\
     \mathcal{C}_{\mu\nu,\lambda}^{(j)}(\boldsymbol{\alpha})  &=  \sum_{\omega=0}^{N_m} \sum_{i=1}^{N_U}  \mathcal{V}_{\omega\mu\nu} b_\omega^{(i)}(\boldsymbol{\alpha}) \langle \tilde{\psi}_{\mu,\lambda}^{(j)} | u^{(i)}_\omega(r)| \phi_{\nu}\rangle\delta_{\nu\lambda}.
    \end{aligned}
\end{equation}

Using the EIM the radial integrals of Eq.~\eqref{eq:petrov-galerkin} are now independent of $\boldsymbol{\alpha}$ and so only need to be computed once in the \textit{offline stage}. The \textit{online} computational cost of solving the equations for a new set of parameters $\boldsymbol{\alpha}$ now involves two steps:
\begin{enumerate}
    \item For each $\omega$, compute the coefficient vector $b_\omega^{(i)}(\boldsymbol{\alpha})$ by multiplying the inverse of the $N_U\times N_U$ matrix $(\boldsymbol{U}_\omega^{\text{EIM}})^{-1}$ with the evaluated potential vector $\boldsymbol{c}_\omega^{\text{EIM}}$, as defined in Eqs.~\eqref{eq:EIM_matrix_form_a} and \eqref{eq:EIM_matrix_form}. Then, reconstruct the radial form factor by multiplying the resulting EIM coefficients with the corresponding basis functions as shown in Eq.~\eqref{eq:EIM_coupling_potential_expansion}. 
    \item Solve the linear system involving the $(N_{\psi}\cdot N_c)\times (N_{\psi}\cdot N_c)$ matrix defined in Eq.~\eqref{eq:petrov-galerkin} for the wavefunction coefficients $c^{(i)}_{\mu}(\boldsymbol{\alpha})$.
\end{enumerate}
For a pedagogical implementation of all equations, we refer the reader to Appendix~\ref{app:two-state}.


\subsubsection{Emulating Cross Sections}\label{sub:emulating_cross}
When attempting to get cross sections for a coupled-channel system, it is not sufficient to solve for a single total wavefunction $\Psi_\lambda$ as defined in Eq.~\eqref{eq:total_wavfunction_decomp}. Rather, one must solve for all possible $\Psi_\lambda$ allowed per $J_\pi$ with $\lambda=1, 2, ..,N_c$. Physically, this means that one must solve for the wavefunction corresponding to all possible incoming and outgoing boundary conditions in the coupled $J_\pi$ block, including those for which the target is initially in an excited state, even if our physical problem restricts the target to be in the ground state in the incoming channel \cite{herman2007empire}. For high-fidelity solvers using a Numerov implementation such as {\sc fresco} \cite{fresco}, it is often easier to solve for the \textit{fundamental matrix of solutions} $\boldsymbol{Y}(r)$. A second-order differential equation allows for two linearly independent solutions, taking all possible channel combinations yields this $N_c\times N_c$ matrix. From the \textit{fundamental matrix of solutions} one can then construct any radial solution as a linear combination of its columns:
\begin{equation}
    \psi_{\mu,\lambda}(r) = \sum_{\beta=1}^{N_c}Y_{\mu\beta}(r)c_{\beta\lambda}.
\end{equation}
Numerically, $\boldsymbol{Y}$ is obtained by imposing linearly independent boundary conditions such as:
\begin{equation}
    Y_{\mu\beta}(r_{\textit{min}}) = \delta_{\mu\beta}C(l_\beta)
\end{equation}
Where, $C(l_\beta)$ is any suitable normalization function dependent of the orbital angular momentum between the projectile and target and $r_{\text{min}}$ is the minimum radius of numerical integration. From the $\boldsymbol{Y}(r)$ matrix we can then obtain the scattering R-matrix by matching to the asymptotic solutions at some maximum radius $a$ \cite{thompson2009nuclear}:
\begin{equation}
    \boldsymbol{R} = \boldsymbol{Y}(a)[a\boldsymbol{Y}'(a)]^{-1}.
\end{equation}
A simple derivation shows that if $\boldsymbol{\hat{\Psi}}(r)$ is the full matrix of solutions whose rows $\mu$ and columns $\lambda$ correspond to the radial wavefunctions $\psi_{\mu,\lambda}(r)$, then we also have:
\begin{equation}
    \boldsymbol{R} =  \boldsymbol{\hat{\Psi}}(a)[a\boldsymbol{\hat{\Psi}}'(a)]^{-1} .
\end{equation}
This means that the Petrov-Galerkin equations defined in Eq.~\eqref{eq:petrov-galerkin} must be solved for each possible $\lambda=1,2, ..,N_c$ in each $J_\pi$ block independently and with its own basis expansion, leading to $N_c$ solutions per block. This is slightly different from \cite{garcia2023wavefunctionmomentum} where $\frac{N_c(N_c+1)}{2}$ solutions are required. Once the R-matrix has been obtained we can compute the S-matrix:
\begin{equation} \label{eq:S_matrix}
    \boldsymbol{S} = [\boldsymbol{O}(a)-a\boldsymbol{R}\boldsymbol{O}
'(a)]^{-1}[\boldsymbol{I}(a)-a\boldsymbol{R}\boldsymbol{I}'(a)].
\end{equation}
Here $\boldsymbol{I}(\boldsymbol{O})$ are matrices whose diagonals are the free incoming(outgoing) Coulomb functions corresponding to each channel $\nu$ in the coupled $J_\pi$ block.

\section{Implementation}\label{sec:Implementation}

The formulation presented in the previous section is general and applies to a broad class of coupled-channel scattering problems, including projectiles with charge and any type of inelastic couplings. 
As a proof of principle, we implement this formalism for realistic coupled-channel neutron–nucleus inelastic scattering, 
assuming a collective model for the target structure.
In such a case,
the interaction is described by a deformed optical potential composed of three Woods--Saxon (WS) terms. 
In the absence of deformation, the potential takes the form:

\begin{align}\label{Eq: Optical Potential}
      U(r;\boldsymbol{\alpha}) = & \, {-}\Big[V_v f_\text{WS}(r,R_v,a_v) \notag\\
      & \null + iW_v f_\text{WS}(r,R_w,a_w) \notag 
    \Big] \\
      & \null -i4a_dW_d \frac{d}{dr}f_\text{WS}(r,R_d,a_d) ,
\end{align}
where the Woods–Saxon form factor is defined as:
\begin{equation}
    f_\text{WS}(r,R,a) = \bigg[1+\text{exp}\Big( \frac{r-R}{a}\Big)  \bigg]^{-1}.
\end{equation}
Since we do not expect internal excitations of the incident nucleon at the beam energies considered, we neglect the internal Hamiltonian of the projectile. Furthermore, since the focus of this work is the coupled-channel system arising from target excitations, we simplify the projectile and neglect its spin. This amounts to not having a spin-orbit term in the interaction potential, as can be seen in Eq.~\eqref{Eq: Optical Potential}. A generalization to include the spin-orbit term is straightforward. By neglecting the intrinsic spin of the projectile, the resulting channel states \(\nu\) are defined as:
\begin{equation}\label{eq:implementation_couplings}
    |\nu(\boldsymbol{\hat{r}}, \boldsymbol{\xi_t}) \rangle = |(l_\nu I^t_{\nu})J \rangle = \bigg [   i^{l} \boldsymbol{\mathcal{Y}}_{l_\nu}^{m_l}(\boldsymbol{\hat{r}})  \otimes  \boldsymbol{\Phi}_{I_\nu}^t(\boldsymbol{\xi_t}) \bigg ]_{JM} .
\end{equation}
The nucleus' deformation is incorporated in this framework (a rigid rotor) by adding an angular dependence (for example in the form of a spherical harmonic function $\mathcal{Y}$), relative to the body-fixed frame, to the radial coordinate of the optical potential:
\begin{equation}
    V(\boldsymbol{r},\boldsymbol{\xi}_t;\boldsymbol{\alpha})=  U(r-\delta_\omega \mathcal{Y}_\omega^{0} (\boldsymbol{\hat{r}'});\boldsymbol{\hat{\alpha}}).
\end{equation}
Here $\boldsymbol{\hat{r}'}$ is the  $\boldsymbol{\hat{r}}$ vector rotated to the body-fixed frame defined by the Euler angles $\boldsymbol{\xi_t}$.  Therefore, our parameters in this model, $\boldsymbol{\alpha}=\{\delta_\omega,\boldsymbol{\hat{\alpha}}\}$, consist of one deformation parameter and nine WS parameters,
\begin{equation}\label{eq:parameters}
    \begin{aligned}
            \boldsymbol{\alpha} = \{ \delta_\omega, V_v,R_v,a_v,W_v,R_w,a_w,
            W_d,R_d,a_d\}.
    \end{aligned}
\end{equation}
Here, $\delta_\omega$ defines the length and multipole order $\omega$ of the deformation. Following the derivation in \cite{temura} or \cite{thompson2009nuclear}, a simple expansion for small deformations yields:
\begin{align}\label{eq:expansion}
    V(\boldsymbol{r},\boldsymbol{\xi}_t;\boldsymbol{\alpha}) = \sqrt{4\pi}U(r;\boldsymbol{\hat{\alpha}})\mathcal{Y}_0^0(\boldsymbol{\hat{r}'}) - \delta_\omega U'(r;\boldsymbol{\hat{\alpha}}) \mathcal{Y}_\omega^0(\boldsymbol{\hat{r}'}).
\end{align}
The first term in the coupling potential yields the diagonal Woods-Saxon terms, the second are the potentials that generate the coupling. In this framework, the sum over $\omega$ in Eq.~\eqref{eq:cc_final} includes only two terms. We naturally have $f_{\omega=0}(r;\boldsymbol{\alpha})=U(r;\boldsymbol{\alpha})$ and $f_{\omega>0}(r;\boldsymbol{\alpha})=\delta_\omega U'(r;\boldsymbol{\hat{\alpha}})$. 

All details of the coupling matrix elements defined in Eq.~\eqref{eq:couplings_term}, applied to this system can be found in Appendix A. In the next two subsections, we will discuss two implementations for two coupling orders $\omega$ of the above formalism. Specifically, the coupling of the $0^+$ ground state to a $2^+$ or $3^-$ excited target state. 


The speed gain of the emulator is strongly implementation dependent, but it can be understood as arising from two distinct sources. The first gain comes from restructuring the calculation to avoid redundant work: quantities such as the Petrov--Galerkin integrals (Eq.~\eqref{eq:petrov-galerkin}), coupling matrix elements (Eq.~\eqref{eq:couplings_term}), Coulomb functions (Eq.~\eqref{eq:S_matrix}), and other system-specific terms that do not change across different parameter sets $\boldsymbol{\alpha}$ can be precomputed once in the \textit{offline stage} and reused. The second gain comes from the reduced dimensionality of the problem: by projecting onto the reduced-basis functions, the emulator replaces the full system with a much smaller linear system (Eq.~\eqref{eq:petrov-galerkin}), which can be solved far more efficiently than using a finite-differences method such as Runge-Kutta or Numerov.

Large speed-ups of more than three orders of magnitude have been reported using dimensionality-reduction techniques \cite{garcia2023wave}. 
In neutron--nucleon scattering, the main efficiency gain comes from reducing the basis size: a standard $R$-matrix approach---commonly used to solve the scattering equations \cite{descouvemont2010r}---typically requires $\sim 10^2$ Lagrange--Legendre basis functions \cite{baye2015lagrange}, whereas their RBM (eigenvector continuation) needs only $\sim 8$ RBM functions. Since the computational cost of a matrix inversion scales as $\mathcal{O}(N_\psi^3)$, this reduction directly accounts for the dramatic speed-up observed. In our application to low-energy neutron--nucleus scattering, however, we find that a single RBM function corresponds to approximately $2$--$2.5$ Lagrange--Legendre $R$-matrix functions. As a result, the improvement is more modest, though still significant.

The implementation presented here combines {\sc fresco} for generating wavefunctions with the \texttt{JAX} library (on CPU) for efficient vectorization of the Petrov--Galerkin equations (Eq.~\ref{eq:petrov-galerkin}). Unlike a conventional Numerov solver, the linear algebra formulation of the equations naturally enables the construction of an emulator that can exploit GPU parallelization. This advantage, however, comes at the cost of additional memory requirements for storing the precomputed quantities discussed in the previous section. For a fair comparison with {\sc fresco}, we restrict the results shown here to the CPU-based emulator. All calculations---both emulator and {\sc fresco}---are performed on a laptop. The {\sc fresco} results correspond to the fastest converged calculation, i.e., the configuration with the coarsest integration grid that yields stable results. To minimize overhead when comparing to {\sc fresco}, only the minimal standard output was requested in {\sc fresco} (no wavefunctions or auxiliary quantities were printed). The reported {\sc fresco} timing therefore includes only the actual computation and standard output plus the time required to load the cross sections into a \texttt{Python} array.

\section{Results}\label{sec:Results}

\subsection{\texorpdfstring{$^{48}$Ca$(n,n')^{48}$Ca$(2^+)$}{48Ca(n,n'}48Ca(2+)}


\begin{figure}[ht!]
    \centering
    \includegraphics[width=0.95\linewidth]{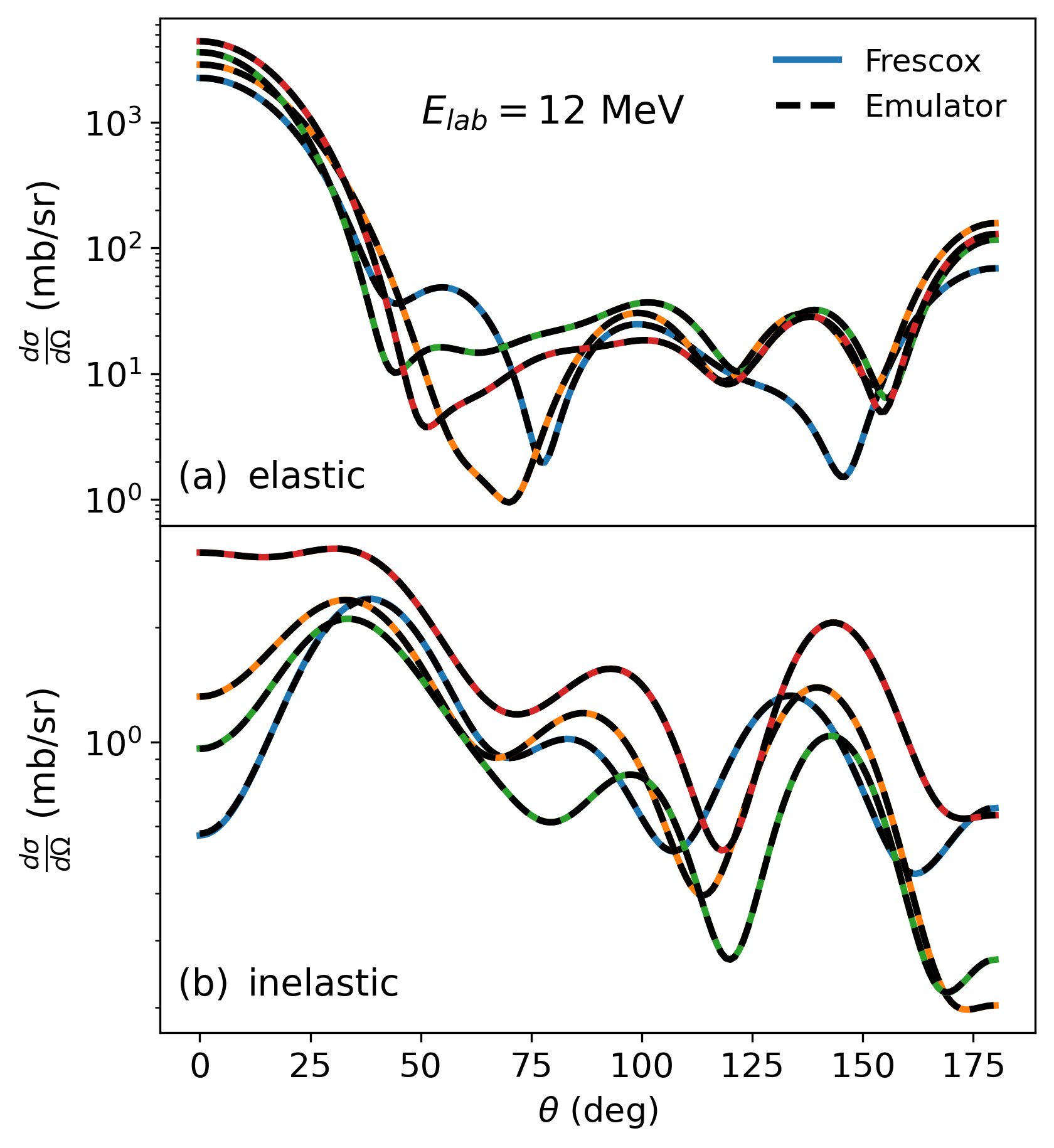}
    \includegraphics[width=0.95\linewidth]{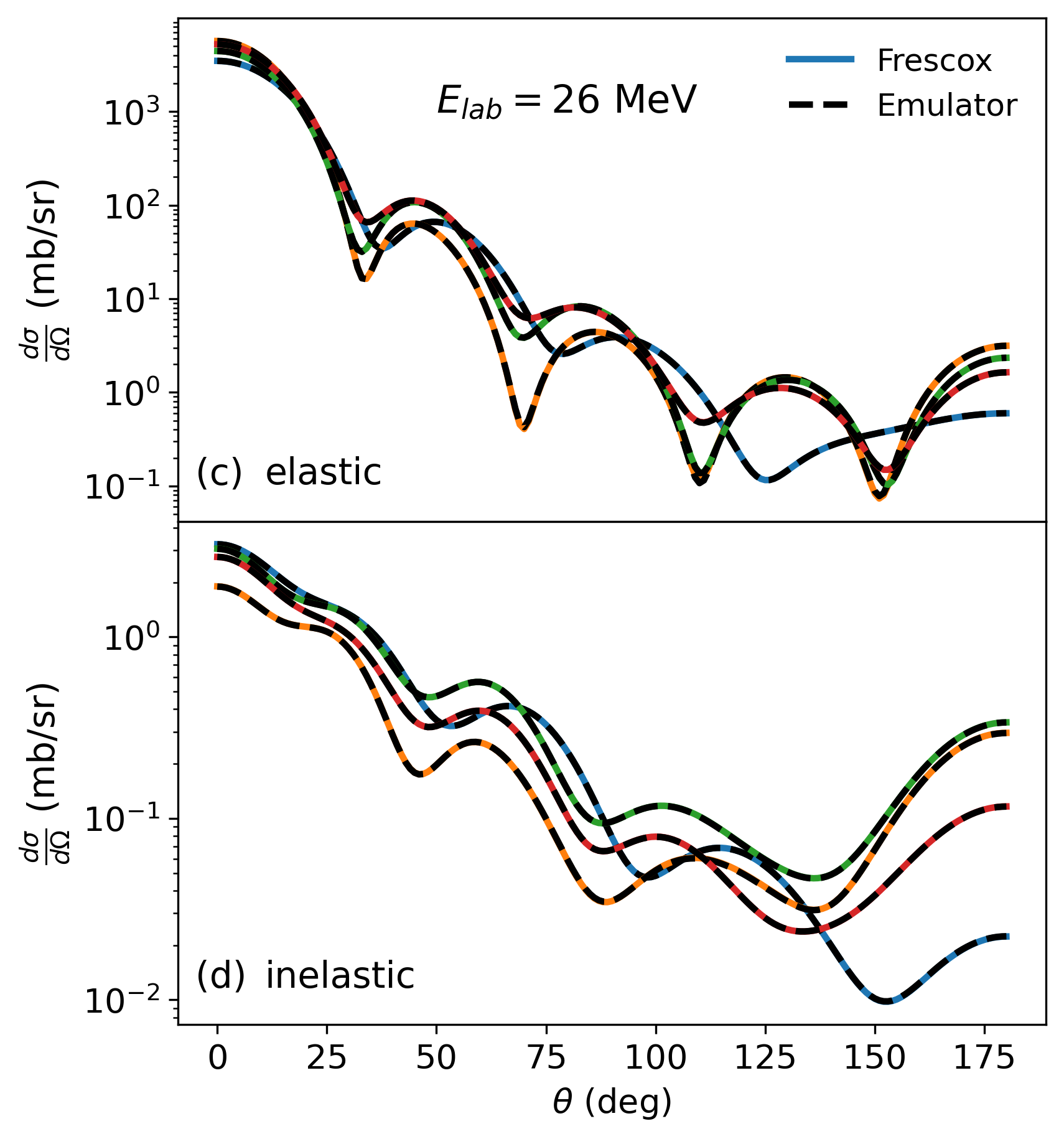}
    \caption{Calculated differential cross sections for $^{48}$Ca$(n,n')^{48}$Ca inelastic scattering at two incident energies. Panels (a)–(b) show the elastic and inelastic cross sections at $E_\text{lab}=12$\,MeV, while panels (c)–(d) show the corresponding results at $E_\text{lab}=26$\,MeV. Solid colored lines denote the {\sc fresco} calculations; dashed black lines denote the emulator results with $N_\psi=12$ and $N_U=12$. Each curve corresponds to a different choice of interaction parameters $\boldsymbol{\alpha}$.
}
    \label{fig:48Ca_cross_emulated}
\end{figure}

We first consider the case of neutron scattering on $^{48}$Ca with excitation of the $2^+$ vibrational state at 3.83 MeV. This is a canonical benchmark for coupled-channel methods, since the $2^+$ state is a textbook example of vibrational excitation. In our framework, this corresponds to setting $\omega=2$ in the deformation expansion of Eq.~\eqref{eq:expansion}. 

We choose two energies, $E_\text{lab}=12$ MeV and $E_\text{lab}=26$ MeV, corresponding to different energy regimes and different signatures in the cross-section diffraction pattern. Using {\sc fresco}, we find that including 15 partial waves is sufficient for these calculations, that is $J_\text{max}=15$. To test the accuracy of the emulator, we select physically relevant parameter values. We adopt the Koning--Delaroche (KD) global parametrization~\cite{koning2003local} as a reference. For the quadrupole deformation parameter of this system we use $\beta_2=0.107$ \cite{48Ca_beta_2}, from which the deformation length, $\delta_2$, can be easily computed.
At each incident energy, the KD and previously mentioned $\beta_2$  parameters serve as the central values around which the sampling is performed. 
Training and testing parameter sets are generated using the \texttt{latinhypercube()} sampling routine from the \texttt{SciPy} library, 
with parameter ranges extending up to $\pm 20\%$ of their corresponding central value.



We find that at least 200 training \textit{snapshots} are required to obtain well-converged RBM and EIM basis functions. 
In this work, 300 \textit{snapshots} are used for the training of all systems considered, and an additional 50 points are selected for testing. 
Note that although the samples for the EIM and RBM are drawn from parameter boxes of the same size, they are generated independently. 
Figure~\ref{fig:SVD_components} presents the first four principal components obtained from the wavefunction decomposition of the coupled-channel emulator at the two incident energies. 
The elastic channels ($l=0$, $I=0^+$) are shown in panels~(a) and~(c), while the inelastic channels ($l=2$, $I=2^+$) are shown in panels~(b) and~(d). 
The corresponding singular values of the RBM basis, consistent with the findings of Ref.~\cite{odell2024rose}, indicate that only a small number of components are sufficient to capture the dominant channel dynamics, thereby supporting the dimensional-reduction strategy employed in the emulator.

One favorable feature of the RBM emulator is the ability to tune the trade-off between speed and accuracy by varying the number of EIM ($N_{\text{U}}$) and RBM ($N_{\psi}$) basis functions included in the approximation. 
Figure~\ref{fig:48Ca_cross_emulated} compares the high-fidelity cross sections produced by {\sc fresco} with those generated by the emulator using $N_{\psi}=12$ and $N_{\text{U}}=12$. Even though the cross sections span several orders of magnitude with complex angular dependence, the RBM emulator is able to reliably predict both elastic and inelastic observables at different energy regimes. 

\begin{figure}
    \centering
    \includegraphics[width=\linewidth]{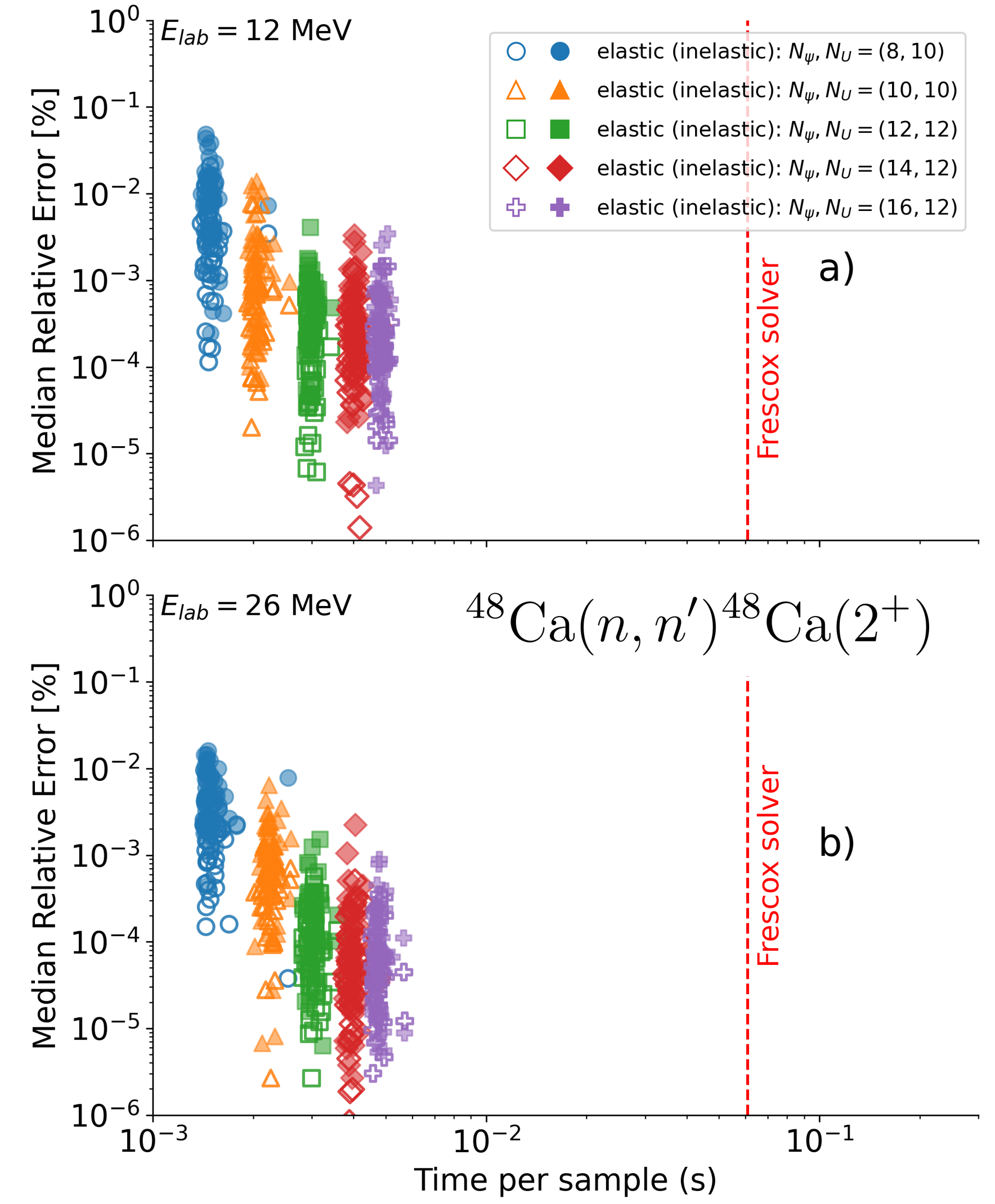}
    \caption{Computational Accuracy vs.\ Time (CAT) plot illustrating the trade-off between accuracy and computational speed for both the CC emulator and the high-fidelity {\sc fresco} solver in the calculation of the differential cross section for the elastic (open markers) and inelastic (filled markers) channels of the 
$^{48}$Ca$(n,n')$$^{48}$Ca$(2^+)$ reaction at $E_{\text{lab}} = 12$~MeV [panel~(a)] and $E_{\text{lab}} = 26$~MeV [panel~(b)]. 
The horizontal axis represents the evaluation time per calculation, while the vertical axis shows the accuracy, defined as the median relative error in the differential cross section with respect to the {\sc fresco} result, computed across 50 test parameter sets centered on the corresponding KD values. 
For the CC emulator, the number of basis functions in the wavefunction expansion ($N_{\psi}$) and the EIM representation ($N_{\text{U}}$) were varied between 8 and 16. 
The vertical red dashed line indicates the average {\sc fresco} solver evaluation time of 60.9~ms. 
The emulator achieves relative errors below $10^{-1}$ while providing speedups approaching one and a half orders of magnitude, depending on the basis size employed.}
    \label{fig:CAT_coupled_channels_48Ca}
\end{figure}

The resulting trade-off between speed and accuracy is quantified in Fig.~\ref{fig:CAT_coupled_channels_48Ca}, which presents the Computational Accuracy vs.\ Time (CAT) performance of the emulator relative to the high-fidelity {\sc fresco} solver. In the CAT plot we vary the emulator basis sizes between $N_\psi = 8-16$ and $N_{\text{U}}=10-12$. The red dashed line corresponds to the average time it takes to run ${\sc fresco}$. For $^{48}$Ca$(n,n')^{48}$Ca$(2^+)$, it corresponds to 60.9 milliseconds. The ${\sc fresco}$ calculation is run in the same processor as the emulator.
Across the 50 test parameter values at $E_{\text{lab}} = 12$ and $26$~MeV, the emulator achieves median relative errors in the cross section well below $10\%$ while reducing the evaluation time by almost two orders of magnitude. 
Increasing the basis dimensions $N_{\psi}$ and $N_{\text{U}}$ systematically improves the accuracy while retaining a substantial computational advantage. 
These results demonstrate that the coupled-channel emulator efficiently and reliably reproduces both elastic and inelastic observables for the $^{48}$Ca system.



\subsection{\texorpdfstring{$^{208}$Pb$(n,n')^{208}$Pb$(3^-)$}{208Pb(n,n')208Pb(3-)}}

As a second case study, we examine neutron scattering on $^{208}$Pb, including excitation of the low-lying collective $3^-$ state at 2.61~MeV. 
The $^{208}$Pb system provides a stringent benchmark for the emulator due to its heavy mass, strong absorption, and the significance of octupole couplings in the low-energy spectrum. 
In the formalism of Eq.~\eqref{eq:parameters}, this corresponds to a deformation of multipolarity $\omega = 3$, with coupling potentials proportional to the derivative of the Woods--Saxon interaction weighted by $Y_3^0(\hat{r}')$. 
Calculations are performed at the same two incident energies, $E_{\text{lab}} = 12$ and $26$~MeV, including partial waves up to $J_{\text{max}} = 15$. 
As before, the KD parametrization at each energy serves as the reference point for sampling. The octupole deformation parameter, also being sampled, was centered around $\beta_3=0.0375$ \cite{beta_3_lead}.

\begin{figure}
    \centering
    \includegraphics[width=\linewidth]{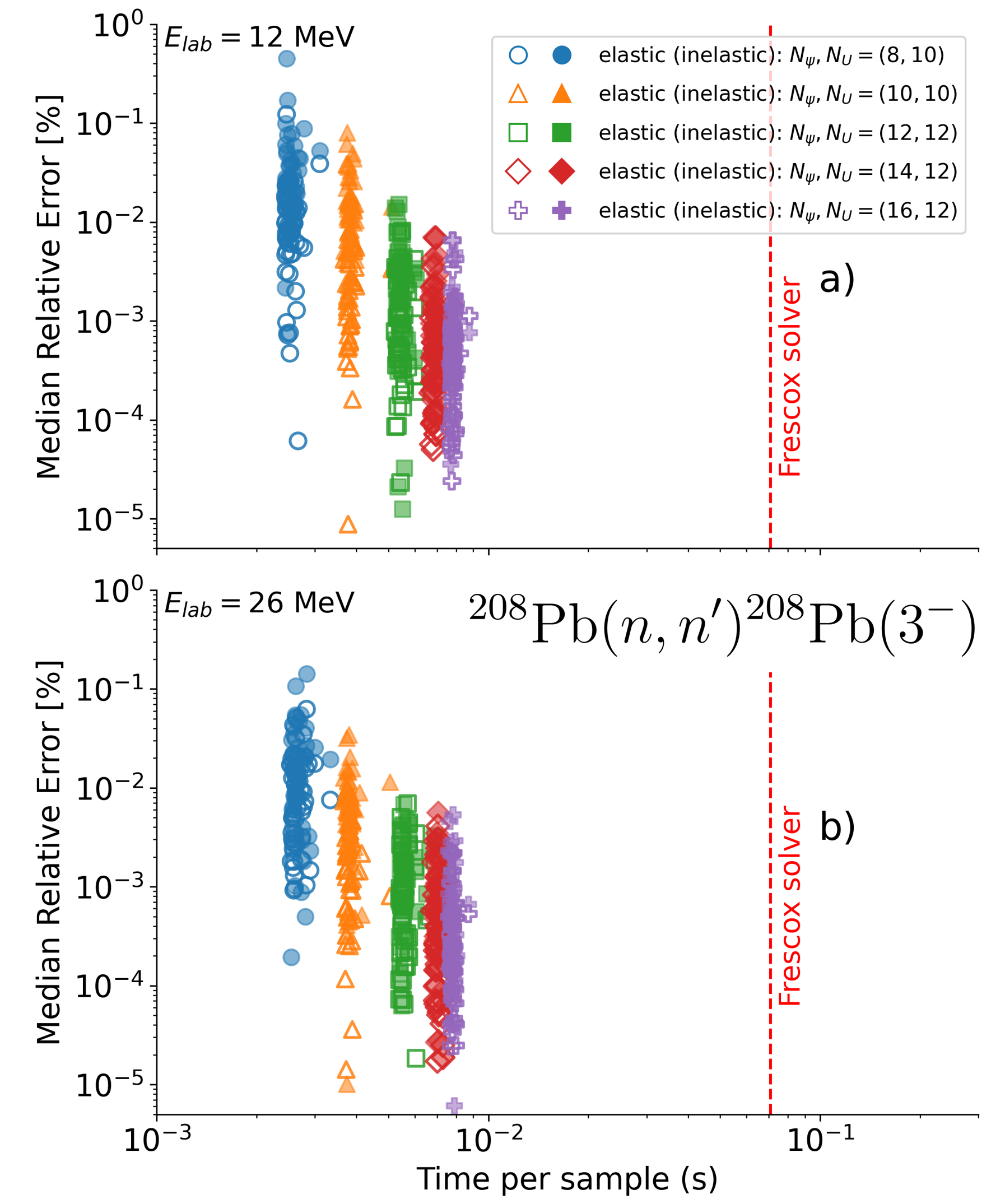}
    \caption{Computational Accuracy vs.\ Time (CAT) plot illustrating the trade-off between accuracy and computational speed for both the coupled-channel (CC) emulator and the high-fidelity {\sc fresco} solver in the calculation of the differential cross section for the elastic (open markers) and inelastic (filled markers) channels of the 
$^{208}$Pb$(n,n')$$^{208}$Pb$(3^-)$ reaction at $E_{\text{lab}} = 12$~MeV [panel~(a)] and $E_{\text{lab}} = 26$~MeV [panel~(b)]. 
The horizontal axis represents the evaluation time per calculation, while the vertical axis shows the accuracy, defined as the median relative error in the differential cross section with respect to the {\sc fresco} result, computed across 50 test parameter sets centered on the corresponding KD values. 
For the CC emulator, the number of basis functions in the wavefunction expansion ($N_{\psi}$) and the EIM representation ($N_{\text{U}}$) were varied between 8 and 16. 
The vertical red dashed line indicates the {\sc fresco} solver evaluation time of 75.1~ms. 
The emulator achieves relative errors below $10^{-1}$ while providing speedups around one and a half orders of magnitude, depending on the basis size employed.}
    \label{fig:CAT_coupled_channels_208Pb}
\end{figure}

Figure~\ref{fig:CAT_coupled_channels_208Pb} presents the CAT plots for both energies, with $E_{\text{lab}} = 12$~MeV shown in the top panel and $E_{\text{lab}} = 26$~MeV in the bottom panel. 
As in the calcium case, the emulator basis sizes were varied between $N_{\psi} = 8$--$16$ and $N_{\text{U}} = 10$--$12$, while the {\sc fresco} solver result, indicated by the red dashed line, corresponds to an evaluation time of 75.1~ms. 
The emulator consistently achieves median relative errors well below $10\%$ across the test parameter sets, confirming its reliability in describing both elastic and inelastic observables in a heavy, strongly coupled system with octupole couplings.

The robustness of the emulator across different multipole orders and mass regimes underscores its potential as a versatile tool for large-scale studies of coupled-channel reactions. Nevertheless, we observe a marginal decrease in the speed-up compared to the quadrupole case. This reduction stems from the need to emulate the wavefunctions for all incoming and outgoing boundary conditions, as discussed in Subsection~\ref{sub:emulating_cross}. To obtain the Petrov--Galerkin coefficients in Eq.~\eqref{eq:petrov-galerkin}, one must solve a linear system of dimension $(N_c \cdot N_\psi)\times (N_c \cdot N_\psi)$, leading to a computational complexity of $\mathcal{O}(N_c^{3} N_\psi^{3})$. 

Since this calculation must be repeated $N_c$ times, corresponding to the possible incoming boundary conditions, the total cost of emulating cross sections scales as $\mathcal{O}(N_c^{4} N_\psi^{3})$. For example, octupole couplings can connect up to five different partial waves by angular-momentum selection rules, in contrast to quadrupole couplings which involve at most four. Other coupled-channel emulators based on reduced-basis methods appear to face similar scaling limitations \cite{garcia2023wavefunctionmomentum,Hagino2025}. In practice, however, for the medium- and heavy-mass systems studied here, this trade-off still yields speed-ups exceeding one and a half orders of magnitude relative to the high-fidelity solver, while maintaining median cross-section errors below the percent level. This demonstrates that, although unfavorable scaling can diminish efficiency gains in large coupled systems, the emulator framework remains robust and highly efficient in the low-energy neutron-nucleus cases studied.

\section{Conclusions and Outlook}\label{sec:Conclusions}

In this work we have developed and demonstrated a coupled-channel emulator for neutron--nucleus scattering based on a reduced-basis Petrov--Galerkin formalism for non-affinely parametrized potentials and realistic nuclear couplings (rigid-rotor). The emulator combines precomputation of system-dependent quantities with a dimensionality-reduced representation of the scattering wavefunctions, enabling fast and accurate evaluation of differential cross sections. We used the benchmark cases of $^{48}$Ca$(n,n')^{48}$Ca$(2^+)$ and $^{208}$Pb$(n,n')^{208}$Pb$(3^-)$ at $E_\text{lab}=12$ MeV and $E_\text{lab}=26$ MeV. The method generalizes robustly from medium-mass to heavy nuclei and across multipole couplings, energies and masses.

The observed efficiency gains are primarily the result of two factors: (i) restructuring the computation to avoid redundant evaluation of integrals, coupling matrix elements, and Coulomb functions, and (ii) reducing the dimensionality of the linear system through projection onto the reduced-basis functions. Although unfavorable scaling with the number of coupled channels $N_c$ introduces some reduction in speed-up for large coupled systems (e.g., octupole excitations),
the emulator remains highly efficient and accurate within the regimes relevant to low-energy nucleon--nucleus reactions.

The coupled-channel emulator developed here offers a practical and flexible tool for accelerating reaction theory, with controlled trade-offs of speed and accuracy in a way that makes systematic studies of complex nuclear systems tractable. Its demonstrated performance in medium- and heavy-mass nuclei motivates future work extending the approach to broader classes of couplings and embedding it within modern uncertainty quantification pipelines.

Further work must be carried out to understand the limitations of this approach and if indeed the poor scaling with increasing number of coupled-channels can be overcome. 
For the moment,
the emulator appears to not be appropriate for 
nuclear reactions including 
many states such as those in \cite{Soukhovitski2016}. 
%
Looking ahead, several extensions of this framework are promising. 
The use of GPU acceleration, already natural in the linear-algebra formulation, offers additional gains in speed that can further expand the scope of feasible parameter studies.



\section*{Acknowledgements}

This research was supported by the CSSI program Award OAC-2004601 (BAND collaboration).
F.M.N. and K.B. acknowledge  support of the U.S. Department of Energy Grant No. DE-SC0021422.
R.J.F. acknowledges support from the National Science Foundation Award Nos.\ PHY-2209442/PHY-2514765 and the NUCLEI SciDAC program under award DE-FG02-96ER40963.

\clearpage

\appendix

\section{Details for coupled-channel equations and observables}\label{app:scattering_details}

\subsection{Coupling Matrix Elements}

The effective n-target potential $V(\boldsymbol{r},\boldsymbol{\xi_t})$ for a deformed rigid rotor can be decomposed in multipoles $\omega$:
\begin{equation}\label{eq:full_potential_decomposition}
    \begin{aligned}
    V(\boldsymbol{r},\boldsymbol{\xi_t};\boldsymbol{\alpha}) &= \sqrt{4\pi}\sum_{\omega m_\omega} U_\omega (r;\boldsymbol{\alpha})D^{\omega}_{m_\omega 0}\mathcal{Y}^{m_\omega}_\omega(\boldsymbol{\hat{r}})\\&
    = \sqrt{4\pi} \sum_\omega U_\omega(r;\boldsymbol{\alpha})P_\omega(\cos \theta'),
    \end{aligned}
\end{equation}
where $D$ is the Wigner rotation matrix relating the lab-frame vector $\boldsymbol{\hat{r}}$ to the body-fixed frame $\boldsymbol{\hat{r'}}$, often referred to as the Euler angles. The exact form the radial dependence of these multipole couplings can be derived from the properties of Legendre polynomials:
\begin{equation}
    U_{\omega}(r;\boldsymbol{\alpha}) = \frac{1}{\sqrt{4\pi}}\int_0^{2\pi} V(\boldsymbol{r,\xi_t};\boldsymbol{\alpha})P_\omega(\cos\theta')\sin\theta'd\theta'd\phi' \;.
\end{equation}
For small deformation lengths $\delta_\omega$, such as those used in this work, one can perform a first order expansion which simplifies Eq.~\eqref{eq:full_potential_decomposition}:
\begin{equation}
    \begin{aligned}
    V(\boldsymbol{r},\boldsymbol{\xi_t};\boldsymbol{\alpha}) &\approx U(r;\boldsymbol{\hat{\alpha}}) - \sum_\omega\delta_\omega U'(r;\boldsymbol{\hat{\alpha}}) P_\omega(\cos\theta')
    \end{aligned}
\end{equation}
The eigenstates of a deformed rotor are given by:
\begin{equation}
    \boldsymbol{\Phi_{I_\nu}}(\boldsymbol{\xi_t}) = \sqrt{\frac{2I_\nu+1}{8\pi^2}} D^{I_\nu\quad*}_{MK_\nu}(\boldsymbol{\xi_t}).
\end{equation}
We can now calculate the \textit{geometric} part of the coupling potential defined in Eq.~\eqref{eq:couplings_term}. As in Sect.~\ref{sec:Implementation}, we neglect the internal structure of the projectile, as well as its spin. Following the derivation in \cite{thompson2009nuclear} we project onto a state:
\begin{equation}
    |\nu(\boldsymbol{\hat{r}}, \boldsymbol{\xi_t}) \rangle = |(l_\nu I^t_{\nu})J \rangle,
\end{equation}
and obtain the following matrix elements:
\begin{equation}\label{eq:geometric_coupling_expansion}
    \begin{aligned}
        \mathcal{V}_{\omega\mu\nu}
       & = \langle (l_\mu I_{\mu}^{t} )J'|| P_\omega(\cos\theta') || (l_\nu I_\nu^t)J \rangle  \\
       & = \delta_{JJ'}(-1)^{\omega+J+l_\nu+ I_\mu} \\
       & \times i^{l_\nu-l_\mu} \hat{I}_\nu\hat{l}_\nu\hat{\omega}
        \begin{Bmatrix}
        l_\nu & I_\nu & J \\
        I_\mu & l_\mu & \omega
        \end{Bmatrix}\\
        & \times
        \langle l_\nu0, \omega 0 | l_\mu 0\rangle \langle I_\nu K_\nu, \Omega 0|I_\mu K_\mu\rangle,
    \end{aligned}
\end{equation}

\subsection{Calculating Inelastic Cross Sections}
Once the S-matrix is obtained (see Eq.~\eqref{eq:S_matrix}), computing the cross sections is straightforward. In the most general nucleus-nucleon scattering case, where we include projectile spin, the cross section for populating a target in state $I_\nu$ from some initial state $I_0$ can be expressed as:
\begin{equation}
    \frac{d\sigma_{I_\nu}}{d\Omega} = \frac{1}{(2s+1)(2I_0+1)}\sum_{m_{s'}m_{I_\nu} m_s m_{I_0}} |f_{m_{s'}m_{I_\nu} m_s m_{I_0}} (\theta)|^2
\end{equation}
In the (l--s) coupling basis, the scattering amplitude is:
\begin{equation}
\begin{aligned}
f_{m_{s'}m_{I_\nu} m_s m_{I_0}}(\theta)
&=\delta_{m_{s'}m_s} \delta_{m_{I_\nu} m_{I_0}}f_{C,I_0}(\theta)
\\
&\times \frac{4\pi}{2i}
   \sum_{\substack{JM\\ ljI_0,\, l'j'I_\nu\\ mm',\, m_j,\, m_{j'}}}
   Y_{l'm'}(\theta,0)\,Y_{l0}^*(0,0) \\
&\quad\times \langle l'm's'm_{s'}|j'm_{j'}\rangle
   \langle j'm_{j'}I_\nu m_{I_\nu}|JM\rangle \\
&\quad\times \langle JM|jm_{j}I_0 m_{I_0}\rangle
   \langle jm_{j}|l0sm_{s}\rangle e^{i\sigma_{l' I_\nu}} \\
&\quad\times   \left(\mathbf{S}^{J}_{l'j'I_\nu,\, ljI_0}
   - \delta_{ll'}\delta_{jj'}\delta_{I_0 I_\nu}\right)
   \frac{e^{i\sigma_{l I_0}}}{k_{\nu}},
\end{aligned}
\end{equation}
where $f_{C,I_0}$ is the amplitude from the Coulomb interaction alone, $\sigma$ is the Coulomb phase shift and $k_{\nu}$ is the momentum corresponding of a channel labeled by $\nu$ \cite{thompson2009nuclear}.

\section{2-level system example}\label{app:two-state}
In this appendix, we present a pedagogical example illustrating the formalism developed in Sec.~\ref{sec:formalism}.
We consider a simple and general two-level system and explicitly derive the corresponding Petrov–Galerkin equations.
Starting from Eq.~\eqref{eq:cc_before_expansion}, we consider the transition from an initial channel $\lambda$ to a final state $\nu$, resisticting only to two channels. This leads to a pair of coupled equations to be solved for each incoming boundary condition corresponding to $\lambda = 1, 2$. We work out explicitly the equation when the initial channel is $\lambda=1$. We have:
\begin{equation}\label{Eq:two_coupled}
    \begin{aligned}
    \big( T_{1}+V_{11}-(E - \epsilon_1 \big )\psi_{1,1} & = -V_{12}\psi_{2,1} \\
    \big( T_{2}+V_{22}-(E - \epsilon_2) \big) \psi_{2,1} & = -V_{21}\psi_{1,1}
    \end{aligned}
\end{equation}
Although not shown explicitly, $\psi$ and $V$  depend on the radial coordinate $r$ and the interaction parameters $\boldsymbol{\alpha}$. To construct the reduced basis elements, we obtain $N_s$ high-fidelity solutions to the coupled equations ~\eqref{Eq:two_coupled} and perform the PCA. The set of $N_\psi$ chosen basis elements for the two level system with incoming wave $\lambda=1$ are then expressed as:
\begin{equation}
    \begin{aligned}
    \{ \tilde{\psi}^{(k)}_{1,1}(r) \}_{k=1}^{N_\psi} &= \text{PCA}\bigg [ \{  \psi_{1,1}(r;\boldsymbol{\alpha_i}) - \phi_{1
}(r)    \}_{i=1}^{N_s}\bigg] \\
\{ \tilde{\psi}^{(k)}_{2,1}(r) \}_{k=1}^{N_\psi} &= \text{PCA}\bigg [ \{  \psi_{2,1}(r;\boldsymbol{\alpha_i})   \}_{i=1}^{N_s}\bigg] \;,
    \end{aligned}
\end{equation}
where $\phi_1$ is the free solution.
Following Sect.~\ref{sec:formalism}, we then approximate the solution to each wavefunction in the coupled system as a linear combination of the corresponding basis elements $\tilde{\psi}$ and some coefficients $c$:
\begin{equation}
    \begin{aligned}
        \psi_{1,1}(r;\boldsymbol{\alpha}) &\approx 
     \phi_{1} (r) + \sum_{k=1}^{N_{\psi}} c^{(k)}_{1}(\boldsymbol{\alpha}) \tilde{\psi}^{(k)}_{1,1}(r)\\
     \psi_{2,1}(r;\boldsymbol{\alpha}) &\approx 
      \sum_{k=1}^{N_{\psi}} c^{(k)}_{2}(\boldsymbol{\alpha}) \tilde{\psi}^{(k)}_{2,1}(r)
    \end{aligned}
\end{equation}
To obtain the coefficients we now perform the Petrov-Galerkin projection used in Eq.~\eqref{eq:petrov-galerkin-def}:
\begin{equation}
    \begin{aligned}
        \mathcal{D}^{(jk)}_{1,1} &= \langle \tilde{\psi}_{1,1}^{(j)} | T_1 (r)- (E-\epsilon_1 ) | \tilde{\psi}_{1,1}^{(k)}\rangle,\\
        \mathcal{D}^{(jk)}_{2,1} &= \langle \tilde{\psi}_{2,1}^{(j)} | T_2 (r)- (E-\epsilon_2 ) | \tilde{\psi}_{2,1}^{(k)}\rangle.
    \end{aligned} 
\end{equation}
These terms involve only the kinetic operator and the channel energy. Next we  calculate the terms pertaining to the interaction:
\begin{equation}
    \begin{aligned}
      \mathcal{U}_{11,1}^{(jk)}(\boldsymbol{\alpha}) &= \sum_{\omega=0}^{N_m} \mathcal{V}_{\omega 11}\langle \tilde{\psi}_{1,1}^{(j)} |f_\omega(r; \boldsymbol{\alpha})|\tilde{\psi}_{1,1}^{(k)} \rangle, \\
      \mathcal{U}_{12,1}^{(jk)}(\boldsymbol{\alpha}) &= \sum_{\omega=0}^{N_m} \mathcal{V}_{\omega 12}\langle \tilde{\psi}_{1,1}^{(j)} |f_\omega(r; \boldsymbol{\alpha})|\tilde{\psi}_{2,1}^{(k)} \rangle, \\
      \mathcal{U}_{21,1}^{(jk)}(\boldsymbol{\alpha}) &= \sum_{\omega=0}^{N_m} \mathcal{V}_{\omega 21}\langle \tilde{\psi}_{2,1}^{(j)} |f_\omega(r; \boldsymbol{\alpha})|\tilde{\psi}_{1,1}^{(k)} \rangle, \\
      \mathcal{U}_{22,1}^{(jk)}(\boldsymbol{\alpha}) &= \sum_{\omega=0}^{N_m} \mathcal{V}_{\omega 22}\langle \tilde{\psi}_{2,1}^{(j)} |f_\omega(r; \boldsymbol{\alpha})|\tilde{\psi}_{2,1}^{(k)} \rangle
    \end{aligned} 
\end{equation}
Note that the sum over $\omega$ often collapses due to angular momentum selection rules.

In the case of a non-affinely parametrized interaction, the integrals above must be replaced by those of Eq.~\eqref{eq:update_eim_coeffs}. The remnant term in the Petrov-Galerkin equations coming from the added free solution yields:
\begin{equation}
    \begin{aligned}
     \mathcal{R}_{1,1}^{(j)} &= \langle \tilde{\psi}_{1,1}^{(j)} |T_1 (r)- (E-\epsilon_1 ) | \phi_{1}\rangle, \\
     \mathcal{C}_{11,1}^{(j)} (\boldsymbol{\alpha})  &=  \sum_{\omega=0}^{N_m} \mathcal{V}_{\omega11}\ \langle \tilde{\psi}_{1,1}^{(j)} | f_\omega (r; \boldsymbol{\alpha})| \phi_{1}\rangle, \\
     \mathcal{C}_{21,1}^{(j)} (\boldsymbol{\alpha})  &=  \sum_{\omega=0}^{N_m} \mathcal{V}_{\omega21}\ \langle \tilde{\psi}_{2,1}^{(j)} | f_\omega (r; \boldsymbol{\alpha})| \phi_{1}\rangle.
    \end{aligned} 
\end{equation}

Now, we cast the Petrov-Galerkin equations into linear algebra form:
\begin{equation}\label{example_galerkin}
\setlength{\arraycolsep}{10pt} 
\renewcommand{\arraystretch}{2} 
    \begin{aligned}
        \begin{pmatrix}
           \mathcal{D}^{(jk)}_{1,1} + \mathcal{U}_{11,1}^{(jk)}(\boldsymbol{\alpha})  &   \mathcal{U}_{12,1}^{(jk)}(\boldsymbol{\alpha})    \\
          \mathcal{U}_{21,1}^{(jk)}(\boldsymbol{\alpha})   &   \mathcal{D}^{(jk)}_{2,1} + \mathcal{U}_{22,1}^{(jk)}(\boldsymbol{\alpha})   \\
        \end{pmatrix}
        \begin{pmatrix}
            c_1^{(k)}\\
            c_2^{(k)}\\
        \end{pmatrix} \\
         = 
        \begin{pmatrix}
            \mathcal{R}_{1,1}^{(j)} + \mathcal{C}_{11,1}^{(j)} (\boldsymbol{\alpha}) \\ 
            \mathcal{C}_{21,1}^{(j)} (\boldsymbol{\alpha}). \\
        \end{pmatrix}
    \end{aligned}
\end{equation}
These equations are now solved for some unknown $c_1$ and $c_2$. In  addition to the solutions for $\lambda=1$ (shown above),
we also need to compute the solutions for the incoming channel $\lambda=2$. The full scattering wavefunction can then be expressed as:
\begin{equation}
    \boldsymbol{\hat{\Psi}} = 
\begin{pmatrix}
\psi_{1,1} & \psi_{1,2} \\
\psi_{2,1} & \psi_{2,2}
\end{pmatrix}.
\end{equation}
From the asymptotic form of this full solution, we extract the full scattering matrix (see Eq.~\eqref{eq:S_matrix}).



\input{output.bbl}

\end{document}

%% file: output.bbl
%